\documentclass[number,preprint,3p]{elsarticle}
\usepackage{titlesec}
\usepackage{color}
\usepackage[dvipsnames]{xcolor}
\usepackage{soul}
\usepackage{graphicx}
\usepackage{subcaption}
\usepackage{algorithm}
\usepackage[colorlinks]{hyperref}
\makeatletter
\gdef\urlauthor#1#2{\g@addto@macro\@elsuads{\let\corref\@gobble%
     \def\@@tmp{#1}\raggedright\eadsep
     {\ttfamily\url{\expandafter\strip@prefix\meaning\@@tmp}}\space(#2)%
     \def\eadsep{\unskip,\space}}%
}
\gdef\emailauthor#1#2{\stepcounter{ead}%
     \g@addto@macro\@elseads{\raggedright%
      \let\corref\@gobble\def\@@tmp{#1}%
      \eadsep{\ttfamily\href{mailto:\expandafter\strip@prefix\meaning\@@tmp}{\expandafter\strip@prefix\meaning\@@tmp}}
      (#2)\def\eadsep{\unskip,\space}}%
}
\makeatother
\usepackage{amssymb}
\usepackage{amsthm}
\usepackage{amsmath}
\usepackage{amssymb}
\usepackage{mathtools}
\usepackage{dsfont}
\usepackage{booktabs}
\usepackage{url}
\usepackage{scrextend}
\usepackage{epstopdf}
\usepackage{float}
\usepackage{tcolorbox}
\usepackage{tablefootnote}
\usepackage[perpage]{footmisc}
\usepackage{lineno}
\def\r{\mathbb{R}}
\def\rn{\mathbb{R}^n}
\def\defi{\coloneqq}
\def\tr{^\intercal}

\newcommand{\vect}[1]{\boldsymbol{#1}}



\makeatletter
\let\@afterindenttrue\@afterindentfalse
\makeatother

\biboptions{numbers,sort&compress,square}
\journal{arXiv}
\usepackage{enumitem}

\setlist[itemize]{itemsep=3pt, topsep=3pt, parsep=3pt, partopsep=3pt}
\setlist[enumerate]{itemsep=3pt, topsep=3pt, parsep=3pt, partopsep=3pt}

\begin{document}
	\begin{frontmatter}
		\renewcommand{\thefootnote}{\fnsymbol{footnote}}
		\title{Physics-based linear regression for high-dimensional \\forward uncertainty quantification}
		\author[1]{Ziqi Wang\corref{cor1}}
         \ead{ziqiwang@berkeley.edu}
         \cortext[cor1]{Corresponding author}
		\address[1]{Department of Civil and Environmental Engineering, University of California, Berkeley, United States}
		\begin{abstract}
We introduce linear regression using physics-based basis functions optimized through the geometry of an inner product space. This method addresses the challenge of surrogate modeling with high-dimensional input, as the physics-based basis functions encode problem-specific information. We demonstrate the method using two proof-of-concept stochastic dynamic examples.
\end{abstract}
  
		
	\end{frontmatter}
	
	\renewcommand{\thefootnote}{\fnsymbol{footnote}}
	
 \section{Problem Statement and Introduction}
Consider an end-to-end computational model $\mathcal{M}:\vect x\in\rn\mapsto y\in\r$ that maps an $n$-dimensional input $\vect{x}$ to a $1$-dimensional output $y$. The input $\vect{x}$ is an outcome of a high-dimensional random vector $\vect{X}$ with a probability measure $\mathbb{P}_{\vect{X}}$ defined on $(\mathbb{R}^n, \mathcal{B}_n)$, where $\mathcal{B}_n$ is the Borel $\sigma$-algebra on $\mathbb{R}^n$. The random output $Y$ is associated with a probability space $(\r,\mathcal{B},\mathbb{P}_{Y})$, where $\mathbb{P}_{Y}$ is the push-forward measure of $\mathbb{P}_{\vect X}$ induced by $\mathcal{M}$. We seek a surrogate model $\hat{\mathcal{M}}:\vect x\in\rn\mapsto \hat y\in\r$ to approximate the statistical properties of $Y$. This task is challenging due to the high dimensionality of $\vect x$. Specifically, conventional data-fitting surrogate models such as polynomial chaos expansion \cite{soize2004physical, xiu2010numerical, torre2019data, novak2024physics} and Gaussian process (Kriging) \cite{kitanidis1997introduction, murphy2012machine, yang2019physics} become increasingly ineffective as the number of input variables increases. Injecting domain/problem-specific prior knowledge into the surrogate modeling process has proven to be a promising approach to mitigate the curse of dimensionality, as reflected in the advancements in scientific machine learning \cite{raissi2019physics, zhu2019physics, psaros2023uncertainty} and multi-fidelity uncertainty quantification \cite{peherstorfer2018survey, wang2024optimized, XIAN2024113069}. This short communication adapts and reformulates the recent works \cite{wang2024optimized, XIAN2024113069} on physics-based surrogate modeling into a simple, unified framework of linear regression. In this linear regression, the basis functions are simplified physics-based models with tuning parameters. An inner product space is introduced to facilitate the training of these physics-based basis functions. Two proof-of-concept examples are presented to demonstrate the proposed approach.

\section{Linear Regression Using Physics-Based Basis Functions}\label{sec:opt}
We represent the surrogate model $\hat{\mathcal{M}}$ by the linear regression:
  \begin{equation}\label{eq:surr}
 \hat{\mathcal{M}}(\vect x)= \vect s(\vect x)\vect w\,,    
  \end{equation}
where $\vect s(\vect x)=[s_1(\vect x),s_2(\vect x),\dots,s_m(\vect x),1]$ is a row vector of basis functions and $\vect w=[w_1,w_2,\dots,w_{m+1}]\tr$ is a column vector of weights. We define $s_i(\vect x)$, $i=1,2,\dots,m$, as physics-based models with tuning parameters $\vect\theta_i$. Note that the basis function vector $\vect s(\vect x)$ contains a constant basis as its last component.


We define the inner product between functions of $\vect x$ as:
  \begin{equation}\label{eq:innerprod}
 \langle f, g\rangle\defi\mathbb{E}_{\vect X}\left[\kappa\left(f(\vect X)\,,g(\vect X)\right) \right]\,,   
\end{equation}
where $f\,,g:\rn\mapsto\r$ are functions of $\vect x$, and $\kappa:\r\times\r\mapsto\r$ is a symmetric and positive definite kernel function that induces an inner product in the function space of $f$ or in an implicit feature space of $\phi(f)$, known as the kernel trick \cite{hofmann2008kernel}. If the kernel is linear, i.e., $\kappa(af+bf',g)=a\kappa(f,g)+b\kappa(f',g)$ for scalars $a$ and $b$, the inner product is defined in the original function space of $f$; otherwise, the inner product is formulated in an implicit feature space induced by the kernel. To ensure a finite inner product, we require $\kappa(f,f)$ be integrable with respect to the measure $\mathbb{P}_{\vect X}$.

Given a training set $\mathcal{D}=\{(\vect x^{(i)},
\mathcal{M}(\vect x^{(i)}))\}_{i=1}^N$, we optimize $s_1(\vect x;\vect\theta_1)$ by solving:
\begin{equation}\label{eq:opt1}    \vect\theta_1^{\star}=\mathop{\arg\max}_{\vect\theta}\dfrac{\left\langle \mathcal{M},s_1(\vect\theta)\right\rangle}{\sqrt{\left\langle \mathcal{M},\mathcal{M}\right\rangle\left\langle s_1(\vect\theta),s_1(\vect\theta)\right\rangle}}\,,
\end{equation}
where the inner product terms can be evaluated by the sample estimates using $\mathcal{D}$. This optimization aims to align $s_1$ with the direction of $\mathcal{M}$, in the inner product space defined by Eq.~\eqref{eq:innerprod}. 

Given $s_1(\vect x;\vect\theta_1^{\star})$, we optimize $s_2(\vect x;\vect\theta_2)$ by solving:
\begin{equation}\label{eq:opt2}
    \vect\theta_{2}^{\star}=\mathop{\arg\max}_{\vect\theta}\dfrac{\left\langle \mathcal{M}-\mathcal{P}_{s_1}(\mathcal{M}),s_2(\vect\theta)\right\rangle}{\sqrt{\langle \mathcal{M}-\mathcal{P}_{s_1}(\mathcal{M}),\mathcal{M}-\mathcal{P}_{s_1}(\mathcal{M})\rangle\left\langle s_2(\vect\theta),s_2(\vect\theta)\right\rangle}}\,,
\end{equation}
where $\mathcal{P}_{s_1}(\mathcal{M})$ denotes the projection of $\mathcal{M}$ onto $s_1$, defined as:
\begin{equation}\label{eq:proj}
\mathcal{P}_{s_1}(\mathcal{M})\defi \dfrac{\langle\mathcal{M},s_1\rangle}{\langle s_1,s_1\rangle}s_1\,.   
\end{equation}
Eq.~\eqref{eq:opt2} aims to align $s_2$ with the direction of the orthogonal residual, $\mathcal{M}-\mathcal{P}_{s_1}(\mathcal{M})$, between $\mathcal{M}$ and $s_1$. 

Subsequently, given $s_1(\vect x;\vect\theta_1^\star),s_2(\vect x;\vect\theta_2^\star),...,s_{j-1}(\vect x;\vect\theta_{j-1}^\star)$, the $j$-th basis function is obtained from:
\begin{equation}\label{eq:optn}
    \vect\theta_{j}^{\star}=\mathop{\arg\max}_{\vect\theta}\dfrac{\left\langle \mathcal{M}-\sum_{k=1}^{j-1}\mathcal{P}_{s_k}(\mathcal{M}),s_j(\vect\theta)\right\rangle}{\sqrt{\langle \mathcal{M}-\sum_{k=1}^{j-1}\mathcal{P}_{s_k}(\mathcal{M}),\mathcal{M}-\sum_{k=1}^{j-1}\mathcal{P}_{s_k}(\mathcal{M})\rangle\left\langle s_j(\vect\theta),s_j(\vect\theta)\right\rangle}}\,,
\end{equation}
where $\mathcal{P}_{s_k}(\mathcal{M})$ denotes the projection of $\mathcal{M}$ onto $s_k$, expressed by replacing $s_1$ in Eq.~\eqref{eq:proj} with $s_k$. 

Provided with the optimized physics-based basis functions and appended by a constant basis, their weights can be computed using the conventional linear regression solution:
\begin{equation}\label{eq:weight}
    \vect w=(\mathcal{S}\tr\mathcal{S})^{-1}\mathcal{S}\tr\vect{\mathcal{Y}}\,,
\end{equation}
where $\mathcal{S}$ is an $N\times (m+1)$ matrix with entries $\mathcal{S}_{ij}=s_{j}(\vect x^{(i)})$, in which $s_{m+1}(\vect x)\equiv1$, and $\vect{\mathcal{Y}}$ is an $N\times 1$ vector with components ${\mathcal{Y}}_i=\mathcal{M}(\vect x^{(i)})$, in which $\vect x^{(i)}$ and  $\mathcal{M}(\vect x^{(i)})$ are from the training set $\mathcal{D}$. A more general formulation for the weights is $\vect w=\mathop{\arg\min}\langle\mathcal{M}(\vect x)-\vect s(\vect x)\vect w,\mathcal{M}(\vect x)-\vect s(\vect x)\vect w\rangle$, which is identical to the linear regression solution if we use a simple linear kernel $\kappa(f,g)=fg$. 
{\color{black}The following additional remarks may be helpful for applications of the proposed method:}
\begin{itemize}
    \item {\color{black} Physics-based basis functions can be adapted from the original model through various ad hoc methods, including linearization, perturbation, homogenization, temporal/spatial coarse-graining, reduced-order modeling, and relaxation of numerical solvers. The general principle is that the basis functions should achieve significant efficiency gains while preserving some key features of the original model. This is where domain knowledge becomes essential. Many scientific and engineering applications have accumulated a rich body of literature on simplified  models. The proposed method offers a straightforward, unified way to leverage these domain-specific simplifications.}
    \item {\color{black}The kernel function determines how similarity is measured between the original and surrogate model predictions and should be selected in alignment with the objectives of surrogate modeling. A linear kernel is preferable when the goal is a relatively uniform match between the original and surrogate predictions. In contrast, nonlinear kernels can be useful when prioritizing accuracy in specific regions. For instance, to assign greater importance to the right tail of the predictions, a nonlinear kernel of the form $(fg)^a$, $a>1$, can be used}. 
    \item The number of physics-based basis functions can be incrementally increased until further additions yield negligible accuracy improvements. {\color{black} Similarly, the number of training points can be determined through cross-validation, where the training set is gradually increased until prediction accuracy on a validation set shows no significant improvement.}
\end{itemize}


\section{Proof-of-Concept Stochastic Dynamic Examples}
\subsection{Duffing Oscillator}
Consider a Duffing oscillator subjected to Gaussian white noise excitation:
\begin{equation}\label{eq:duff}
  {\ddot z}(t)+2\zeta\omega_n{\dot z}(t)+\omega_n^2z(t)+\beta z^3(t)=a(t)\,,
\end{equation}
where $\zeta=0.05$, $\omega_n=10\,\mathrm{rad/s}$, and $\beta=2000\,\mathrm{m}^{-2}\mathrm{s}^{-2}$. The Gaussian white noise $a(t)$ has a unit spectral intensity. For numerical simulations, we set a cutoff angular frequency of $30\pi\,\mathrm{rad/s}$ and represent $a(t)$ by $200$ independent standard Gaussian random variables $\vect X$ weighted by a Fourier series \cite{chatfield2013analysis}. The input of the end-to-end model $\mathcal{M}$ is outcomes of the $200$-dimensional random vector $\vect X$, while the output is the peak absolute displacement $y=\sup_{t\in[0,10]}|z(t)|$ for a duration of $10$ seconds. 

We design the initial basis function $s_{1}$ by generalizing the first-order perturbation of Eq.~\eqref{eq:duff}:
\begin{equation}\label{eq:s1}
    s_1(\vect\theta)=\sup_{t\in[0,10]}\left|h(t;\theta_1,\theta_2)\ast a(t)-\theta_3\,h(t;\theta_1,\theta_2)\ast \left(h(t;\theta_1,\theta_2)\ast a(t)\right)^3\right|\,,
\end{equation}
where ``$\ast$" denotes convolution, $h(t;\theta_1,\theta_2)$ is the impulse response function for a single-degree-of-freedom linear system parameterized by the natural frequency $\theta_1\geq0$ and damping ratio $\theta_2\geq0$, and $\theta_3\geq0$ is another tuning parameter controlling the contribution of the first-order perturbation. The zeroth order perturbation term, $h(t)\ast a(t)$, describes the response of a tunable linear system subjected to $a(t)$. The first-order perturbation term, $h(t)\ast(h(t)\ast a(t))^3$, describes the response of the same linear system subjected to the cube of the zeroth order perturbation term. {\color{black}The first-order perturbation is adopted here due to its computational efficiency and ability to approximate nonlinear behavior. Three parameters are introduced to ensure flexibility in the basis function while avoiding over-parameterization. A more rigorous model selection approach, such as Bayesian model selection, could also be considered}.

The basis functions $s_i$, $i>1$, are modeled by linearization---Eq.~\eqref{eq:s1} with $\theta_3=0$. If $m$ physics-based basis functions are used, the total number of tuning parameters is $2m+1$. Therefore, the surrogate representation is parsimonious for this high-dimensional problem. Two convolutions are required to evaluate $s_1$, while one convolution is needed for each $s_{i}$, $i=2,3,\dots,m$. Thus, $m+1$ convolutions are required to evaluate the basis function vector once, which takes negligible time for a relatively small $m$. 

\begin{figure}[h]
  \centering
  \includegraphics[width=1.0\textwidth] {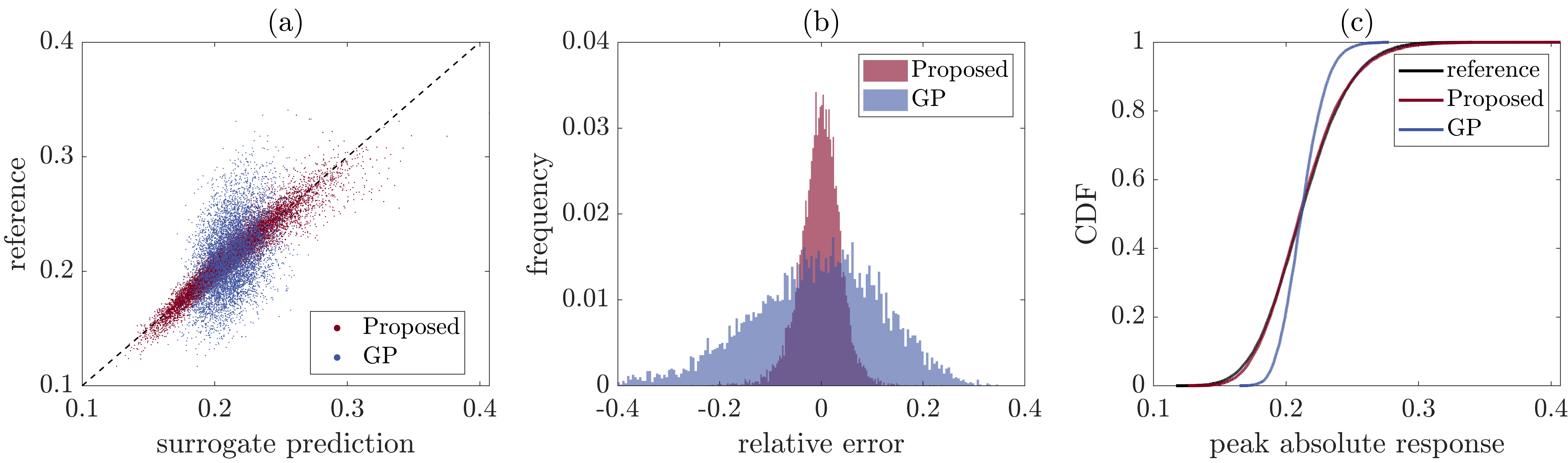}
  \caption{\color{black}\textbf{Performance of the physics-based linear regression for a nonlinear random vibration problem: (a) scatter plot comparing reference solutions with physics-based surrogate model and Gaussian process (GP) predictions, (b) histogram of relative errors, and (c) cumulative distribution function predictions.} The physics-based regression is trained on 30 samples, where the 200-dimensional Gaussian vector $\vect{X}$ characterizing $a(t)$ serves as the input and the peak absolute response as the output. The GP is trained using 3000 samples, where $a(t)$ discretized into 100 time points is the input and the absolute peak response the output. Testing various covariance and trend functions for the GP revealed no significant improvements in prediction accuracy. The physics-based regression exhibits a mean relative error of $-0.3\%$, a first quartile of $-2.5\%$, and a third quartile of $2.3\%$. In contrast, the data-fitting GP has a mean relative error of $-1.4\%$, a first quartile of $9.7\%$, and a third quartile of $8.0\%$.}
  \label{fig:1}
\end{figure}

We optimize the physics-based basis functions using the formulas in Section \ref{sec:opt} and a training set of only $30$ samples from random realizations of the white noise and their corresponding peak absolute responses. A simple linear kernel  $\kappa(f,g)=fg$ is used to define the inner product. An alternative is $\kappa(f,g)=(f-\mathbb{E}_{\vect X}[f(\vect X)])(g-\mathbb{E}_{\vect X}[g(\vect X)])$, which is also linear. We did not find significant evidence to favor one option over the other. The number of physics-based basis functions is adaptively determined, such that by increasing the basis, the reduction in the mean square relative training error is less than 1\%. This number typically ranges from $2$ to $4$ due to stochastic variations in the training set. Using a test set of $10^4$ random samples, Figure \ref{fig:1} compares the reference Runge–Kutta–Fehlberg solutions with the surrogate model predictions. {\color{black}The predictions of the Gaussian process regression, trained on  $3000$ samples of $a(t)$, are also shown for comparison, serving as a baseline for pure data-fitting surrogates.} 


{\color{black}
\subsection{One-Dimensional Heterogeneous Diffusion}
Consider the one-dimensional pressure diffusion equation:
\begin{equation}
    \dfrac{\partial p(z,t)}{\partial t} = \dfrac{\partial}{\partial z}\left(D(z) \dfrac{\partial p(z,t)}{\partial z}\right),
\end{equation}
where $z \in [0, 1]\,\mathrm{m}$ and the diffusivity $D(z)$ is modeled as a lognormal process with a mean of $10^{-2}\,\mathrm{m^2\,s^{-1}}$, a coefficient of variation of $100\%$, and a squared exponential covariance function with a correlation length of $0.01\,\mathrm{m}$. The injection pressure $p_\text{inj} = p(0, \tau)$ for $\tau > 0$ is assumed to be a constant, uniformly distributed within $[2 \times 10^5, 3 \times 10^5]\,\mathrm{Pa}$. The initial pressure $p_{0} = p(z, 0)$ is also a constant, uniformly distributed within $[1 \times 10^5, 2 \times 10^5]\,\mathrm{Pa}$. A Dirichlet boundary condition $p(1, t) = p_0$ is applied. For numerical simulations, we discretize the random field $D(z)$ into $100$ Gaussian random variables. In conjunction with the random injection and initial pressures, the input to the end-to-end model $\mathcal{M}$ has a dimension of $102$. We consider the total volume flowing out of the system within the period $t \in [0, 10]\,\mathrm{s}$ as the quantity of interest, i.e., the output of $\mathcal{M}$. In our parameter setting, the system is highly unlikely to reach a steady state, making the problem nontrivial.

To approximate $\mathcal{M}$ using physics-based linear regression, we employ homogenization to formulate the basis functions. By homogenizing $D(z)$ into an effective diffusivity $D_\text{eff}$, the diffusion equation can be solved analytically. The closed-form solution for the flow-out volume then serves as an efficient basis function, expressed as:
\begin{equation}\label{eq:ex2basis}
    s_i(\vect\theta) = \frac{(p_\text{inj} - p_0) T D_\text{eff}(\vect\theta)}{L} + 2 (p_\text{inj} - p_0) L \sum_{k=1}^\infty \frac{(-1)^k \left(1 - \exp\left(-D_\text{eff}(\vect\theta) \left(\frac{k\pi}{L}\right)^2 T\right)\right)}{k^2 \pi^2},
\end{equation}
where $L = 1\,\mathrm{m}$, $T = 10\,\mathrm{s}$, and $\vect\theta$ are tuning parameters that determine how $D_\text{eff}$ is derived. Building on the widely used harmonic mean $\langle D^{-1} \rangle^{-1}$ for homogenization, we define $D_\text{eff}$ as:
\begin{equation}
    D_\text{eff}^{-1}(\vect\theta) = \frac{\int_0^L \left|\frac{z}{L} - \theta_2\right|^{\theta_1} D^{-1}(z) \, dz}{\int_0^L \left|\frac{z}{L} - \theta_2\right|^{2\theta_1} \, dz},
\end{equation}
where $\theta_1 \in [0, 10]$ and $\theta_2 \in [0, 1]$. This formula reduces to the harmonic mean when $\theta_1 = 0$. In practice, we truncate the summation in Eq.~\eqref{eq:ex2basis} to the first $50$ terms, as higher-order terms contribute negligibly to the flow-out volume. We use the same formula for all physics-based basis functions; they vary only in their parameters.

We train the physics-based basis functions using $50$ samples of $(D(z), p_\text{inj}, p_0)$ and their corresponding flow-out volumes, computed via the finite difference method. Similar to the previous example, the number of basis functions is determined adaptively, typically varying between $3$ and $7$. Figure \ref{fig:2} illustrates the accuracy of the physics-based regression model using a test set of $10^4$ samples.

\begin{figure}[h]
  \centering
  \includegraphics[width=1.0\textwidth] {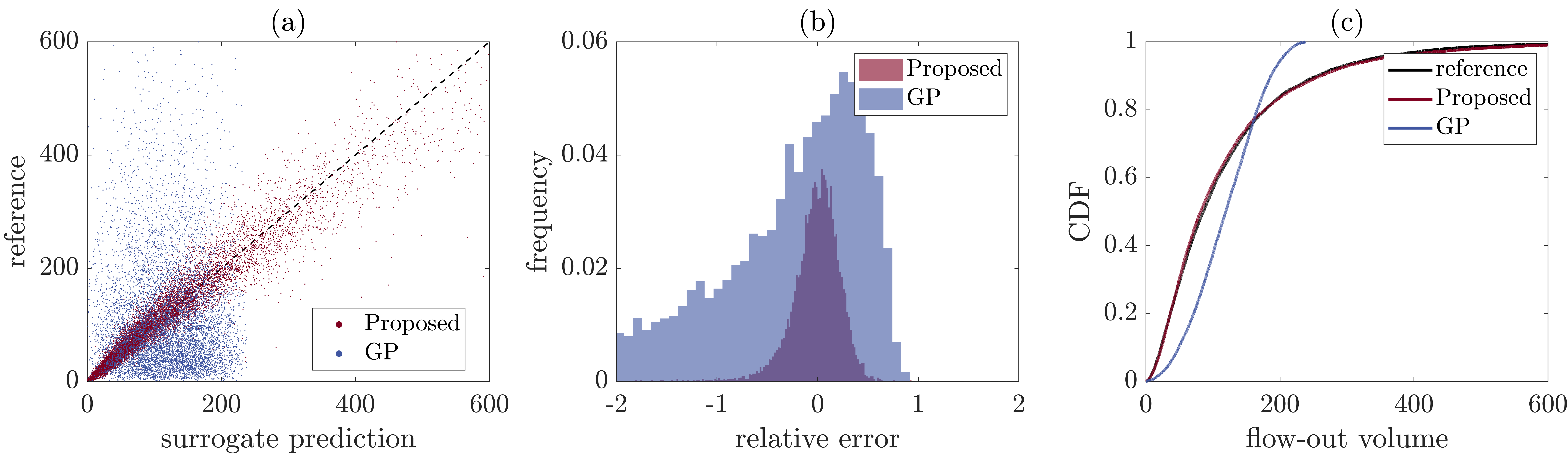}
  \caption{\color{black}\textbf{Performance of the physics-based linear regression for a stochastic heterogeneous diffusion problem: (a) scatter plot comparing reference solutions with physics-based surrogate model and Gaussian process (GP) predictions, (b) histogram of relative errors, and (c) cumulative distribution function predictions.} The physics-based regression is trained on 50 samples, while the GP is trained on 3000 samples. The physics-based regression exhibits a mean relative error of $-2.0\%$, a first quartile of $-13.5\%$, and a third quartile of $13.8\%$. In contrast, the data-fitting GP shows a mean relative error of $-84.7\%$, a first quartile of $-118.5\%$, and a third quartile of $23.9\%$.}
  \label{fig:2}
\end{figure}

\textcolor{black}{The two proof-of-concept examples unanimously suggest that a straightforward data-fitting regression is unlikely to effectively address complex problems characterized by stochastic processes and random fields. In contrast, our proposed method prioritizes specialization over generality, allowing it to tackle the challenge of high dimensionality. Although its ad hoc nature may introduce certain limitations, it also offers significant advantages and opportunities for creating interpretable and efficient surrogate models tailored to specific scientific computing applications.}

}
\section{Conclusion}
This short communication introduces linear regression using physics-based basis functions. The main contribution is to standardize the procedure for training physics-based basis functions. {\color{black}Two proof-of-concept stochastic dynamic examples demonstrate the potential of this approach}. Future research directions may include extensions to multivariate output models, adaptations for rare-event probability estimations, and applications in inverse uncertainty quantification, sensitivity, and optimization problems.

\bibliography{Ref}

 \end{document}